\pdfoutput = 1

\documentclass[12pt,a4paper,final]{article}

\usepackage[backref,
            natbib
            ,style = numeric-comp
            ,maxnames = 4
            ,backend = bibtex
            ,sorting=none
            ]{biblatex}

\usepackage[T1]{fontenc} 
\usepackage[english]{babel} 
\usepackage{csquotes}
\usepackage{amsmath,amssymb,amsfonts,amsthm} 
\usepackage[table,dvipsnames*,svgnames]{xcolor}
\usepackage{fixltx2e}

\usepackage[a4paper,text={490pt,650pt},centering]{geometry}
\usepackage[explicit]{titlesec}
\usepackage[active]{srcltx}
\usepackage{extarrows}

\titleformat{\section}[block]{\normalfont\bfseries\filcenter}{\scshape\thesection}{0.5em}{#1\\\titlerule}
\titlespacing{\section}{5pc}{*3}{*2}[5pc]
\titleformat{\subsection}[runin]{\normalfont\bfseries\filcenter}{\scshape\thesubsection.}{0.5em}{#1}[.---]
\titlespacing{\subsection}{\parindent}{1.5ex plus 0.1ex minus 0.2ex}{0pt}
\titleformat{\subsubsection}[runin]{\normalfont\bfseries\filcenter}{\scshape\thesubsubsection.}{0.5em}{#1}[.---]
\titlespacing{\subsubsection}{1ex plus 0.1ex minus 0.2ex}{1.5ex plus 0.1ex minus 0.2ex}{0pt}

\makeatletter

\let\@authors\@empty
\let\@email\@empty
\let\@affiliation\@empty
\let\@pdfsubject\@empty
\let\@keywords\@empty
\let\@preprint\@empty

\providecommand{\pdfsubject}[1]{\gdef\@pdfsubject{#1}}
\providecommand{\keywords}[1]{\gdef\@keywords{#1}}
\renewcommand{\author}[1]{\ifx\@authors\@empty\toks@\expandafter{#1}\else\toks@\expandafter{\@authors, #1}\fi\edef\@authors{\the\toks@}}
\providecommand{\email}[1]{\ifx\@email\@empty\toks@\expandafter{#1}\else\toks@\expandafter{\@email, #1}\fi\edef\@email{\the\toks@}}
\providecommand{\affiliation}[1]{\gdef\@affiliation{#1}}
\providecommand{\preprint}[1]{\gdef\@preprint{#1}}

\makeatother

\usepackage[hyperindex,breaklinks]{hyperref}

\usepackage{graphicx}

\allowdisplaybreaks[3]
\DeclareMathOperator{\Li}{Li}
\renewcommand{\[}{\begin{equation}}
\renewcommand{\]}{\end{equation}}

\newcommand{\nln}{\notag\\}
\DeclareMathOperator{\sv}{sv}

\bibliography{soft.bib}

\begin{document}

\title{A Note on Soft Factors for Closed String Scattering}

\author{Burkhard U.\ W.\ Schwab}
 \email{\href{mailto:burkhard\_schwab@brown.edu}{burkhard\_schwab@brown.edu}}

\affiliation{%
Brown University\\Department of Physics\\182 Hope St, Providence, RI, 02912%
}

\keywords{Gauge theory, Cluster algebras}
\pdfsubject{Supercluster algebras}
\preprint{Brown-HET-1660}


\makeatletter
\thispagestyle{empty}

\begin{flushright}
\begingroup\ttfamily\@preprint\par\endgroup
\end{flushright}

\begin{centering}
\begingroup\Large\normalfont\bfseries\@title\par\endgroup
\vspace{1cm}

\begingroup\@authors\par\endgroup
\vspace{5mm}

\begingroup\itshape\@affiliation1\par\endgroup
\vspace{3mm}

\begingroup\ttfamily\@email\par\endgroup
\vspace{0.25cm}

\begin{minipage}{17cm}
 \begin{abstract}
In this note, it is shown that closed string graviton scattering amplitudes obey the same subleading soft limit as field theory graviton scattering amplitudes. The result is derived using a combination of recent results and methods including the subleading soft expansion for type I open string gluon disk scattering amplitudes, the single value projection for multiple $\zeta$ values, and the KLT relations in field and string theory.
 \end{abstract}
\end{minipage}
\vspace{1cm}

\end{centering}

\makeatother


\makeatletter
\hypersetup{pdftitle={\@title}}%
\hypersetup{pdfsubject={\@pdfsubject}}%
\hypersetup{pdfkeywords={\@keywords}}%
\hypersetup{pdfauthor={\@authors}}%
\makeatother


\section{Introduction and Discussion}
\label{sec:introduction}

\citeauthor{Cachazo:2014fwa}'s suggestion of a new soft theorem for graviton scattering sparked a surge of interest in the low energy behaviour of these amplitudes \cite{Cachazo:2014fwa}.  Originally treated to leading order by Weinberg \cite{Weinberg:1965nx,Weinberg:1964ew}, it was shown by \citeauthor{Gross:1968in} that there are universal subleading contributions \cite{Gross:1968in}. \citeauthor{Low:1958sn} showed that there is a very similar looking soft expansion for Yang Mills theory \cite{Low:1958sn}. 

The new soft theorem is usually stated as follows. The $N$ graviton tree level amplitude $M_{N}$ factorises in the presence of a soft particle with $k_{N}=q\to0$ (and polarisation tensor $\epsilon_{\mu\nu}$) into a soft part and the $(N-1)$ point amplitude
\[M_{N} = \left(S^{(0)}_{\rm g} + S^{(1)}_{\rm g} + S^{(2)}_{\rm g}\right)M_{N-1} + \mathcal{O}(q).\]  The colour ordered $N$ gluon Yang Mills tree level amplitude $A_{N}$ behaves in a very similar way
\[A_{N} = \left(S^{(0)}_{\rm YM} + S^{(1)}_{\rm YM}\right)A_{N-1} + \mathcal{O}(q).\label{eq:softyang}\] For graviton scattering, when taking the soft particle to be the $N$\textsuperscript{th} particle, the soft factors are
\[S^{(0)}_{\rm g} = \sum_{i=1}^{N-1}\frac{\epsilon_{\mu\nu}p^{\mu}_{i}p^{\nu}_{i}}{q.p_{i}},\qquad S^{(1)}_{\rm g} = \sum_{i=1}^{N-1}\frac{\epsilon_{\mu\nu}p^{\mu}_{i}(q_{\rho}J^{\rho \nu}_{i})}{q.p_{i}},\qquad S^{(2)}_{\rm g} = \sum_{i=1}^{N-1}\frac{\epsilon_{\mu\nu}(q_{\lambda}J^{\lambda \mu}_{i})(q_{\rho}J^{\rho \nu}_{i})}{q.p_{i}}\label{eq:softgrav}\] while the analogous factors for Yang Mills theory are
\[S^{(0)}_{\rm YM} = \frac{\epsilon_{\mu}p^{\mu}_{1}}{q.p_{1}} - \frac{\epsilon_{\mu}p^{\mu}_{n}}{q.p_{n}},\qquad S^{(1)}_{\rm YM} =\frac{\epsilon_{\mu}q_{\nu}J^{\mu\nu}_{1}}{q.p_{1}} - \frac{\epsilon_{\mu}q_{\nu}J^{\mu\nu}_{n}}{q.p_{n}}.\] 

The operator $J_{i}^{\mu\nu}$ denotes the total angular momentum operator of the $i$\textsuperscript{th} particle. $S^{(1)}_{\rm g}$, $S^{(2)}_{\rm g}$, and $S^{(1)}_{\rm YM}$ are therefore derivative operators as opposed to the leading order soft gluon and graviton factors, which are only multiplication operators. The leading and subleading soft graviton factors as well as the soft gluon factors are universal and the soft graviton factor is protected from quantum corrections. 

At approximately the same time when these ideas were first developed (in fact, three years before), the group of symmetries at null infinity of asymptotically flat space was investigated by Bondi, van der Burg, Metzner, and Sachs \cite{Bondi:1962px,Sachs:1962wk}. This symmetry group is known as the BMS group \[{\rm BMS} = T \ltimes SL(2,\mathbb{C})\] and consists of the semidirect product of the Abelian group $T$ of ``supertranslations'' (which are in no way related with supersymmetry) and the group $SL(2,\mathbb{C})$ of non-singular transformations of the two sphere at infinity. The latter can be extended to also contain singular transformations -- called superrotations -- to form a Virasoro algebra \cite{Barnich:2013axa,Barnich:2011mi,Barnich:2009se,Barnich:2011ct}. This line of thought has only recently been put forward by \citeauthor{Barnich:2011ct}. While Low already connected the soft gluon factors with gauge symmetry, it took fifty years for the development of a logical connection between BMS and the soft graviton factor. Recently, such a relation has been presented in a series of papers \cite{He:2014laa,Kapec:2014opa,Strominger2013,Strominger:2013jfa}. Note that the work by \citeauthor{He:2014laa} not only connects the supertranslations with Weinberg's soft factor, but the subsequent work by \citeauthor{Kapec:2014opa} also makes use of the superrotations of the extended BMS group and connect these with the subleading term. Thus the gravitational soft factors are protected by symmetries of the (classical) gravitational $\mathcal{S}$ matrix. The work by \citeauthor{Cachazo:2014fwa} mentioned before \cite{Cachazo:2014fwa} also introduced a novel sub-subleading term  -- $S^{(2)}_{\rm g}$ in \eqref{eq:softgrav} -- which seems universal but is not thought to be protected by any symmetry.

This connection has seen many checks and there have been notable extensions for subleading soft theorems. The latter have been discussed by \citeauthor{Casali:2014xpa} for Yang Mills theory \cite{Casali:2014xpa} in the more modern language of the spinor helicity formalism following \cite{Cachazo:2014fwa}. \citeauthor{Bern:2014oka} as well as \citeauthor{He:2014bga} have shown that the subleading soft factor in Yang Mills theory receives corrections and so does the gravitational subleading soft factor at the loop level \cite{He:2014bga,Bern:2014oka,Bianchi:2014gla}, an alternative prescription has been offered by \citeauthor{Cachazo:2014dia} in \cite{Cachazo:2014dia} which however clashes with the conventional way the soft theorem is understood and used (see also \cite{Bonocore:2014wua}). The universality of the subleading and subsubleading soft factors for arbitrary dimensions in gravity and Yang Mills theory have been checked in \cite{Schwab:2014xua,Afkhami-Jeddi:2014fia,Zlotnikov:2014sva,Kalousios:2014uva}. \citeauthor{Du:2014eca} have shown that there is a nontrivial connection between the Yang Mills subleading soft factor and the gravity subleading soft factor using the field theory KLT relations \cite{Du:2014eca}. These relations impose a novel nontrivial constraint on the field theory KLT kernel. \citeauthor{Broedel:2014fsa} as well as \citeauthor{Bern:2014vva}  \cite{Broedel:2014bza,Broedel:2014fsa,Bern:2014vva} have shown that the form of the soft factors is highly constrained (see also \cite{Larkoski:2014hta}). A derivation of the symmetry principle in ambitwistor string theory has been given by \citeauthor{Geyer:2014lca}. A theory at null infinity describing the observed characteristics of graviton scattering in the soft limit was given by \citeauthor{Adamo2014} in \cite{Adamo2014,Geyer:2014lca}. There have been multiple results on supersymmetric theories as well as QCD \cite{Liu:2014vva,Rao:2014zaa,Luo:2014wea}. On the other hand, a novel extension of BMS has been put forward in \cite{Campiglia:2014yka}.

In essence this note should be treated as an appendix to \cite{Schwab:2014fia}. In this paper it was shown that single trace type I superstring gluon scattering amplitudes satisfy the same subleading soft theorem as field theory Yang Mills scattering amplitudes. As a tree level result, it is independent of the dimension, chosen compactification or the amount of supersymmetry. In the last section of the paper, a way to use the string theory Kawai-Lewellen-Tye (KLT) relations \cite{Kawai:1985xq} 
\[\mathcal{M} = (-1)^{N-3}\kappa^{N-2}\mathcal{A}^{T}.\mathcal{S}.\mathcal{A}\] was mentioned to produce the closed string version of the soft theorem from the result for open strings. In the last equation, $\mathcal{M}$ are either type I or type two closed string amplitudes, $\mathcal{A}$ is a vector of colour ordered type I open string disk amplitudes, and $\mathcal{S}$ is the string theory phase matrix. The notation which will be made more explicit in the text.  Proving the soft graviton theorem for string theory from this perspective involves an additional step. The phase matrix $\mathcal{S}$, which is necessary to combine two open string amplitudes into a closed string amplitude in the KLT prescription must be shown to behave correctly under the soft limit, too. 

In the meantime, the above mentioned result by \citeauthor{Du:2014eca} became available. Using this result it is possible to take a field theory path to the string theory result. To do so, another string theory result connecting type I open string scattering amplitudes with heterotic string scattering amplitudes and closed string scattering amplitudes \cite{Stieberger:2014hba} due to \citeauthor{Stieberger:2014hba} has to be used. However, ref.~\cite{Du:2014eca} only showed the four dimensional case, so here a version which is independent of the number of dimensions is presented as a side product.  It will turn out that combining these three results will make for a rather elegant proof of the subleading soft graviton theorem for closed string scattering amplitudes.

It will only be shown that the subleading soft factor $S^{(1)}_{\rm g}$ can be recovered in this way. A field theory proof that the subsubleading contribution in gravity can be obtained from KLT and the Yang Mills soft theorem is still pending. A moment of reflection will make clear that the subsubleading term in gravity will depend on subsubleading expressions from Yang Mills amplitudes which are not thought to be universal. A possible way to attack this problem would be to use the expansion for MHV amplitudes given in \cite{He:2014bga} to show that the subsubleading term in gravity can be recovered from KLT at least in the MHV case. 

This note's organisation is as follows. In sec.~\ref{sec:preliminaries}, the preliminaries will be presented.  A review of \cite{Schwab:2014fia} is given in ssec.~\ref{sec:soft-behaviour-open}. Ssec.~\ref{sec:single-value-proj} provides an introduction to the single value projection which connects type I gluon scattering amplitudes with heterotic string gluon scattering amplitudes \cite{Brown:2013gia,Stieberger:2014hba}. A byproduct will be the subleading soft theorem for the heterotic string. The derivation of the theorem for gravity from Yang Mills theory via the KLT relations in field theory is presented in \ref{sec:grav-soft-fact}. In this subsection, an extension of the derivation to any dimension is given. The derivation of the subleading soft theorem for closed string graviton scattering, which is very short, is presented in sec.~\ref{sec:closed-soft}. 

\section{Preliminaries}
\label{sec:preliminaries}

The following three subsections provide background for the closed string graviton result and extend some results of previous papers.

\subsection{Soft behaviour of open string disk amplitudes}
\label{sec:soft-behaviour-open}

In \cite{Schwab:2014fia} it was shown that the open string disk amplitudes behave just like the field theory amplitudes in the limit of a single momentum becoming soft. Clearly, this result is reasonable since string theory should not deviate from the field theoretic behaviour in this low energy regime. 

An open string disk amplitude ${\cal A}_{N} = {\cal A}(1,\ldots,N)$ with $N$ particles is given by essentially two pieces. Employing the notation of \cite{Stieberger:2014hba} the first piece is a vector $A$ of colour ordered Yang Mills scattering amplitudes. The entries of this vector are denoted by $A_{\sigma} = A(1,\sigma_{2,N-2},N-1,N)$ where $\sigma\in S_{N-3}$ is a permutation of particles $2$ to $N-2$ \[\sigma: (2,\ldots, N-2) \to (2_{\sigma},\ldots, (N-2)_{\sigma}).\] This vector is dotted into the period matrix $F_{\pi\sigma}$ where $\pi,\sigma \in S_{N-3}$ such that \[{\cal A}_{\pi} =F_{\pi\sigma}A_{\sigma} \label{eq:period}\] where $\pi$ denotes the colour ordering of the string disk amplitude. The form of the period matrix $F$ and can be found in numerous publications, see e.g., \cite{Mafra:2011nv,Mafra:2011nw}. 

In the presence of a soft string momentum $k_{N-2}=q\to0$, here taken to be particle $N-2$, the string disk amplitude factorises
\[{\cal A}(1,\pi(2,\ldots,N-3,q),N-2,N-1) \to (S^{(0)}_{\rm YM} + S^{(1)}_{\rm YM}){\cal A}(1,\pi'(2,\ldots,N-2),N-1,N).\] The numbering of the $N$ particles on the left hand side amplitude was changed for ease of comparison. The $S^{(i)}_{\rm YM}$ are the leading $(i=0)$ and subleading $(i=1)$ gluon soft factors from field theory as in \eqref{eq:softyang}. Note that the Yang Mills soft factor for colour ordered amplitudes always depends on the particles adjacent to the soft particle and only on these. 

\subsection{Single value projection}
\label{sec:single-value-proj}

The next, vital ingredient to introduce is the single value projection $\sv$ \cite{Brown:2013gia,Stieberger:2014hba}. The map $\sv$ is a homomorphism between the Hopf algebra of multiple $\zeta$ values (MZVs) defined by \[\zeta_{n_{1},\ldots,n_{r}} = \sum_{0<k_{1}<\ldots<k_{r}}\prod_{\ell =1}^{r}k_{\ell}^{-n_{\ell}}\] and Brown's single valued multiple $\zeta$ values $\zeta^{\sv}_{n_{1},\ldots,n_{r}}$ defined as the evaluation of single valued (multiple) polylogarithms at $1$. As an example: Single valued (simple) polylogarithms \[D_{m}(z)=\Re\left(i^{m+1}\left[\sum_{k=1}^{m}\frac{(-\log|z|)^{m-k}}{(m-k)!}\Li_{k}(z) - \frac{(-\log|z|)^{m}}{2m!}\right]\right)\label{eq:zagier_ramakrish}\] are a generalisation of the Bloch-Wigner-function 
\[D(z) = \Im(\Li_{2}(z)) + \arg(1-z)\log |z|\] and were studied in slightly different forms by, e.g., Ramakrishnan \cite{MR862642}, Zagier \cite{MR1032949}, Wojtkowiak \cite{MR1051830} and most recently Brown \cite{Brown:2013gia}. In the Ramakrishnan-Zagier form \eqref{eq:zagier_ramakrish} it is $D_{2}(z)=D(z)$. The single-value projection 
\[\sv: \zeta_{n_{1},\ldots,n_{r}}\mapsto \zeta^{\sv}_{n_{1},\ldots,n_{r}}\] satisfies various relationships which derive from the functional equations satisfied by the single-valued polylogarithms. For example it is easy to see that \[\sv(\zeta_{2}) = \sv (\Li_{2}(1))  = D(1) = \zeta^{\sv}_{2} = 0.\] Further examples and relationships can be found in the papers by Brown as well as in \cite{Stieberger:2013wea}. Note that scalars with respect to the Hopf algebra of multiple $\zeta$-values pass through the single value projection.

 The relationship between (colour ordered) type I $N$ gluon disk scattering amplitudes $\mathcal{A}$ and the corresponding heterotic string scattering amplitudes $\mathcal{A}^{\rm HET}$ is governed by $\sv$. In \cite{Stieberger:2014hba} it was shown that these amplitudes are directly connected via the single value projection (and a rescaling $\alpha'\to \alpha'/4$), i.e.,
\[{\cal A}^{\rm HET}_{\pi} = \sv({\cal A}_{\pi}) = \sv(F_{\pi\sigma}) A_{\sigma}.\label{eq:heterotic}\] This is the single value projection of the matrix equation \eqref{eq:period}. The action of $\sv$ on $F_{\pi\sigma}$ is to take the $\zeta$ value expansion of $F_{\pi\sigma}$ and replace every occurrence of a MZV by the corresponding single valued MZV. Since the soft factors are scalar with respect to $\sv$, the relation \[{\cal A}^{\rm HET}_{N+1,\pi} \to \Big(S^{(0)}_{\rm YM}+S^{(1)}_{\rm YM}\Big)_{\pi(q+1),\pi(q-1)}{\cal A}^{\rm HET}_{N,\pi'},\label{eq:hetsoft}\] where $\pi\in S_{N-2}$, $\pi'\in S_{N-3}$, follows almost trivially. As before, $q$ is the soft momentum and the subscript $\pi(q+1),\pi(q-1)$ indicates that the soft factors will depend on the colour ordering in the same way as they depend on the colour ordering in Yang Mills or type I disk scattering amplitudes --- via particles adjacent to the soft particle. Since the field theory limit $\alpha'\to 0$ of $\mathcal{A}^{\rm HET}$ are Yang Mills gluon scattering amplitudes, it is not surprising that the heterotic string scattering amplitudes behave in this way. Despite that, the result will be an important stepping stone for the closed string scattering amplitudes which will be investigated in the next section.

The result in \eqref{eq:heterotic} has another implication \cite{Stieberger:2014hba}. The KLT relations split closed string scattering amplitudes into a right moving and a left moving part. Using \eqref{eq:period} as a basis for the left moving part, let \[\widetilde{{\cal A}}_{N,\sigma} = {\cal A}(1,\sigma(2,\ldots,N-2),N,N-1)\] be a basis for the right moving part.  The KLT relations connect (type I and type II) $N$ graviton string scattering amplitudes with a sum of products of open string amplitudes
\[{\cal M} = (-1)^{N-3}\kappa^{N-2}\sum_{\rho,\sigma\in S_{N-3}}{\cal A}_{\rho}{\cal S}_{\rho,\sigma}\widetilde{{\cal A}}_{\sigma}\] where \cite{BjerrumBohr:2010hn}  \[{\cal S}_{\rho,\sigma} = \prod_{j=2}^{N-2}\sin\left(s_{1,j_{\rho}} + \sum_{k=2}^{j-1}\theta(j_{\rho},k_{\rho}) s_{j_{\rho},k_{\rho}}\right)\] with $\theta(j_{\rho},k_{\rho}) = 1$ if the ordering  of $j_{\rho}$ and $k_{\rho}$ is the same in both $\rho$ and $\sigma$, otherwise it is zero.\footnote{The constant $\kappa$ is the gravitational coupling constant. Note that ${\cal A}$ does not contain any factors of $g_{\rm YM}$.} The scattering amplitudes $\mathcal{A}$, $\widetilde{\mathcal{A}}$, and $\mathcal{M}$ on both sides of the equation have each the same amount of legs $N$. The variables $s_{ij}:= \alpha' p_{i}.p_{j}$ are the usual dimensionless kinematic invariants. Using the $\sv$ projection and some manipulations it is possible to show that closed string graviton scattering amplitudes can be given in terms of only one ``stringy'' factor of ${\cal A}^{\rm HET}$ and contributions of the field theory KLT relations, i.e.,
\[{\cal M}_{N} = (-1)^{N-3}\kappa^{N-2} A^{T}.S_{\rm FT}.{\cal A}^{\rm HET}.\label{eq:KLThet}\] In the last equation, $A^{T}$ is the transpose of the vector of colour ordered Yang Mills scattering amplitudes appearing in \eqref{eq:period}, while $S_{\rm FT}$ is the KLT-kernel \[S_{\rho,\sigma} = \prod_{j=2}^{N-2}\left(s_{1,j_{\rho}} + \sum_{k=2}^{j-1}\theta(j_{\rho},k_{\rho}) s_{j_{\rho},k_{\rho}}\right)\label{eq:kltkernel}\] subject to a basis change of the scattering amplitudes and $s_{ij} = p_{i}.p_{j}$ here. This basis change can be implemented by a matrix $D_{\sigma,\rho}$ which connects the field theory amplitudes $A(1,\sigma(2,\ldots,N-2),N-1,N)$ with the amplitudes $\widetilde{A}(1,\rho(2,\ldots,N-2),N,N-1)$, i.e., $A_{\sigma} = {D_{\sigma,\rho}}\widetilde{A}_{\rho}$ and 
\[S_{\rm FT,\rho,\sigma} = \sum_{\gamma\in S_{n-3}}S_{\rho,\gamma}D_{\gamma,\sigma}.\]

\subsection{Graviton soft factors from KLT}
\label{sec:grav-soft-fact}

The KLT relations have proven to be a very valuable tool for investigations into the classical gravity $\mathcal{S}$ matrix. With the growing understanding of the gauge theory $\mathcal{S}$ matrix the KLT relations give guidance on how to generalise results from the gauge side to gravity. In a recent paper \cite{Du:2014eca}, it was shown that the field theory KLT relations can also be used to prove a connection between the soft factors in gauge theory and gravity.

One important caveat is in order. The results in \cite{Du:2014eca} were derived using methods available only in $d=4$. The KLT relations hold in any dimension, and the subleading soft theorem equally well extends to any dimension \cite{Schwab:2014xua,Afkhami-Jeddi:2014fia,Kalousios:2014uva,Zlotnikov:2014sva}. Trivial considerations then show that the work by \citeauthor{Du:2014eca} can be extended to arbitrary dimensions. In the process of a review of the result, this generalisation will be derived. 
\medskip

The proof in \cite{Du:2014eca} uses an alternative form of the KLT relations as its starting point. Using the notation introduced in ssec.~\ref{sec:single-value-proj}, but now with field theory amplitudes (and stripped gravitational coupling)
\[M_{N} = (-1)^{N-3}\sum_{t=2}^{N-2}\sum_{\sigma,\beta\in S_{N-4}} A_{N}(1,t,\sigma_{2,N-2},N-1,N) S^{p_{1}}_{t,t}S_{\sigma,\beta}^{p_{N-1}}\widetilde{A}_{N}(t,1,N-1,\beta,N)\label{eq:altKLT}.\] This expression is valid in any dimension. Note the appearance of yet another basis $\widetilde{A}$ for the field theory amplitudes on the right hand side. Later on, it will be necessary to use a basis transformation to bring it into the form of \eqref{eq:KLThet}. 

The KLT-kernel $S_{\sigma,\beta}^{p_{N-1}}$ is built around particle $N-1$ which takes the place of particle $1$ as given in \eqref{eq:kltkernel}. More compactly, this can be written as
\[M_{N} = (-1)^{N-3}\sum_{t=2}^{N-2}\sum_{\sigma,\beta\in S_{N-3}} A_{N}(t,\sigma) S^{p_{1}}_{t,t}S_{\sigma,\beta}^{p_{N-1}}\widetilde{A}_{N}(t,\beta).\] A quick calculation will show that the leading contribution to the soft limit follows very naturally from this form.\footnote{The reader will have to excuse the profusion of quantities denoted by the letter $S$. Most of the time, the meaning of the letter should be unambiguous as soft factors always bear a bracketed superscript denoting their level in the soft expansion while KLT kernels always come with subscript indices.} First, send $p_{1}\to 0$. Then $S_{t,t} = p_{1}.p_{t} =: s_{1t}$ and
\begin{align}
A_{N}(1,t,\sigma,N-1,N) &\to (S^{(0)}_{\rm YM} + S^{(1)}_{\rm YM})_{t,N}A_{N-1}(t,\sigma,N-1,N)\\
\widetilde{A}_{N}(t,1,N-1,\beta,N) &\to (S^{(0)}_{\rm YM} + S^{(1)}_{\rm YM})_{N-1,t}A_{N-1}(t,N-1,\sigma,N)
\end{align}
where the subscripts denote the particles adjacent to particle $1$ in the given colour ordering.  After the expansion, the leading order term can be assembled for each $t$ as $S^{(0)}_{t,N}s_{1t}S^{(0)}_{N-1,t}$. This expression multiplies
\[(-1)^{N-4}\sum_{\sigma,\beta\in S_{N-3}} A_{N-1}(t,\sigma,N-1,N)S_{\sigma,\beta}^{N-1}\widetilde{A}_{N-1}(t,N-1,\beta,N)\label{eq:kltrep}\] in the sum over $t$. But \eqref{eq:kltrep} is just another KLT representation of the gravity amplitude $M_{N-1}(2,\ldots,N)$ and therefore in particular independent of $t$. Then, using conservation of momentum it is easy to show that
\[\sum_{t=2}^{N-2} S^{(0)}_{t,N}s_{1t}S^{(0)}_{N-1,t} = -\sum_{t=2}^{N}\frac{\epsilon.p_{t}\epsilon.p_{t}}{s_{1t}} \xlongrightarrow{\epsilon_{\mu}\epsilon_{\nu} \to \epsilon_{\mu\nu}} - S^{(0)}_{\rm g}\] which is Weinberg's soft graviton factor. The minus sign cancels the missing minus sign in \eqref{eq:kltrep} from the definition of $M_{N}$. As usual for KLT in dimensions other than four, the gluon polarisation vectors $\epsilon_{\mu}$ will have to be exchanged for a graviton polarisation tensor $\epsilon_{\mu\nu}$ to get the correct answer.

The subleading term is more subtle, and as was shown in \cite{Du:2014eca}, requires a non-trivial identity for the KLT kernel to hold. The same identity
\[\sum_{t=2}^{N-2}\sum_{\sigma,\beta\in S_{N-4}}D[2,\tilde\sigma,N-1,N|t,\sigma,N-1,N]S_{\sigma,\beta}^{N-1}\Delta_{t} C[t,N-1,\beta,N|2,N-1,\tilde\beta,N] =0 \label{eq:identity}\] where
\[\Delta_{t} = \epsilon_{\mu}p_{1\nu}\left(\frac{\epsilon.p_{t}J^{\mu\nu}_{N-1}}{s_{1,N-1}} + \frac{\epsilon.p_{N}J^{\mu\nu}_{t}}{s_{1,N}} + \frac{s_{1t}\epsilon.p_{N}J^{\mu\nu}_{N-1}}{s_{1,N-1}s_{1,N}} - \frac{\epsilon.p_{t}J^{\mu\nu}_{N}}{s_{1,N}} - \frac{\epsilon.p_{N-1}J^{\mu\nu}_{t}}{s_{1,N-1}} - \frac{s_{1t}\epsilon.p_{N-1}J^{\mu\nu}_{N}}{s_{1,N-1}s_{1,N}}\right)\]
can be shown to arise in the calculation valid for any dimension. The base change matrices $D$ and $C$ are used to transform the basis of colour-ordered Yang-Mills amplitudes via
\[A_{N-1}(t,\tilde\sigma,N-1,N) = \sum_{\sigma\in S_{N-4}}A_{N-1}(s,\tilde\sigma,N-1,N)D[s,\tilde\sigma,N-1,N|t,\sigma,N-1,N]\] where $s$ is arbitrary ($s=2$ in \eqref{eq:identity}) and similarly for $\widetilde{A}_{N-1}$ and $C$.
 The operators $\Delta_{t}$ are more involved in this representation than in the spinor-helicity form given in \cite{Du:2014eca} and depend on the angular momentum operators of particles $N-1$, $N$ and $t$. In this note, no attempts will be made at proving this result. It appears however that the calculation in momentum variables is in fact easier to perform than in spinor helicity variables. Under the above assumption, it can be shown that the subleading order in the soft expansion of the terms in the KLT relation actually lead to the subleading soft graviton factor as given in \eqref{eq:softgrav}, i.e.,
\begin{align}
(-1)^{N-3}\sum_{t=2}^{N-1}\sum_{\sigma,\beta\in S_{N-4}}&\left[\Big(S^{(0)}_{N-1,t}s_{1t}S^{(1)}_{t,n}A_{N-1}\Big)S_{\sigma,\beta}^{N-1}\widetilde{A}_{N-1} + A_{N-1}S_{\sigma,\beta}^{N-1}\Big(S^{(0)}_{N-1,t}s_{1t}S^{(1)}_{t,n}\widetilde{A}_{N-1}\Big)\right]\nln
 &= \sum_{t=2}^{N} \frac{\epsilon.p_{t}\epsilon_{\mu}p_{1\nu}J^{\mu\nu}_{t}}{s_{1t}}M_{N-1} \xlongrightarrow{\epsilon_{m}\epsilon_{\nu}\to\epsilon_{\mu\nu}}S^{(1)}_{\rm g}M_{N-1}
\end{align}

With this, all that is left to do is to assemble the results of this section into the promised proof.

\section{Closed string scattering soft factor}
\label{sec:closed-soft}

This rather short section contains the actual proof of the subleading soft theorem for closed string graviton scattering amplitudes. To perform the final step and show that closed string graviton scattering amplitudes behave like field theory graviton scattering amplitudes in the subleading soft limit, it is only necessary to synthesise the results in ssecs.~\ref{sec:soft-behaviour-open}, \ref{sec:single-value-proj}, and \ref{sec:grav-soft-fact}.

First, revisit equation \eqref{eq:KLThet}. The right hand side can be written more verbosely in the following form 
\[\mathcal{M}_{N} = (-1)^{N-3} \kappa^{N-2} \sum_{\sigma,\beta\in S_{N-3}} A(1,\sigma_{2,N-2},N-1,N)S_{\rm FT}[\sigma|\beta] \sv \mathcal{A}(1,\beta_{2,N-2},N-1,N).\] 
The KLT field theory kernel $S_{\rm FT}$ is related to the KLT kernel $S$ used in sec.~\ref{sec:grav-soft-fact} via a basis transformation which solely depends on combinations of kinematic data $s_{ij}$.  Therefore, this expression can be transformed into the alternative representation of the KLT relations in \eqref{eq:altKLT} using basis transformations which do not interfere with the single value projection, i.e.,
\[\mathcal{M}_{N} = (-1)^{N-3} \kappa^{N-2} \sum_{t=2}^{N-2}\sum_{\sigma,\beta\in S_{N-3}}A_{N}(t,\sigma)S_{t,t}S_{\sigma,\beta}\sv \widetilde{\mathcal{A}}_{N}(t,\beta).\label{eq:altKLThet}\]
This is possible since the heterotic string amplitudes $\mathcal{A}^{\rm HET}$ satisfy the same Kleiss-Kuijf \cite{Kleiss:1988ne} and Bern-Carrasco-Johansson \cite{Bern:2008qj} identities as the field theory amplitudes. Furthermore, the base change matrices connecting right moving $\widetilde{\mathcal{A}}$ and left moving $\mathcal{A}$ are dimensionless \cite{Stieberger:2014hba}. In particular, they do not depend on $\alpha'$. It follows that the leading and subleading soft limit follow in the same way as in sec.~\ref{sec:grav-soft-fact}. Specifically, combining \eqref{eq:hetsoft},\eqref{eq:altKLThet}, and the soft gluon limit for particle $1$
\[ \mathcal{M}_{N} = 
(-1)^{N-3} \kappa^{N-2} \sum_{t=2}^{N-2}\sum_{\sigma,\beta\in S_{N-3}}(S^{(0)} + S^{(1)})_{N,t}A_{N-1}(t,\sigma)s_{1t}S_{\sigma,\beta}(S^{(0)} + S^{(1)})_{t,N-1}\sv \widetilde{\mathcal{A}}_{N-1}(t,\beta)
\]
But the above expression is exactly the same expression as the field theory KLT, which means that all calculations in~\ref{sec:grav-soft-fact} hold in this case. Thus
\[\mathcal{M}_{N}\overset{p_{1}\to 0}{\longrightarrow}\kappa(S^{(0)}_{\rm g} + S^{(1)}_{\rm g})\mathcal{M}_{N-1} + \mathcal{O}(p_{1})\] 
which ends the proof that closed string theory tree level amplitudes feature the same soft limit as their field theory descendants.

In conclusion, all string and field theories connected through KLT relations, the single value projection and the field theory limit $\alpha'\to 0$ must be behaving in the same way under the subleading soft limit at tree level. This encompasses gauge field theory, type I open strings, and heterotic string theory gauge scattering amplitudes. On the gravitational side, it was shown here that the behaviour of (type I, type II, heterotic and bosonic) closed string graviton scattering can be determined from that of open string gluon amplitudes. The field theory limit of these amplitudes coincides with the known gravitational result as it should be. It remains to be seen whether there is an equivalent symmetry statement behind these results as in the field theory case.

\paragraph{Acknowledgements}

The author would like to thank Stephan Stieberger for suggesting the use of the $\sv$ projection. The author is also grateful for discussions with Tim Adamo, Steven Avery, Johannes Br\"odel, Matteo Rosso and Oliver Schlotterer. This work is supported by the US Department of Energy under contract DE-FG02-11ER41742.

\printbibliography
\end{document}